# Antibody-loading of biological nanocarrier vesicles derived from red-blood-cell membranes

*Maryam Sanaee,[1]\* K. Göran Ronquist,[2] Elin Sandberg,[1] Jane M. Morrell,[2] Jerker Widengren,[1] and Katia Gallo[1]*

[1] Department of Applied Physics, School of Engineering Sciences, KTH Royal Institute of Technology, Stockholm 10691, Sweden.

[2] Department of Clinical Sciences, Swedish University of Agricultural Sciences, Uppsala 75007, Sweden

\*Corresponding author: msanaee@kth.se

**ABSTRACT** Antibodies, disruptive potent therapeutic agents against pharmacological targets, face a barrier crossing immune-system and cellular-membranes. To overcome these, various strategies have been explored including shuttling via liposomes or bio-camouflaged nanoparticles. Here, we demonstrate the feasibility to load antibodies into exosome-mimetic nanovesicles derived from human red-blood-cell-membranes. The goat-anti-chicken antibodies are loaded into erythrocyte-membrane derived nanovesicles and their loading yields are characterized and compared with smaller dUTP-cargo. Applying dual-color coincident fluorescence burst methodology, the loading yield of nanocarriers is profiled at single-vesicle level overcoming their size-heterogeneity and achieving a maximum of 38-41% antibody-loading yield at peak radius of 52 nm. The average of 14 % yield and more than two antibodies per vesicle is estimated, comparable to those of dUTP-loaded nanovesicles after additional purification through exosome-spin-column. These results suggest a promising route for enhancing biodistribution and intracellular accessibility for therapeutic antibodies using novel, biocompatible, and low-immunogenicity nanocarriers, suitable for large-scale pharmacological applications.

**KEYWORDS.** Antibody loading, nanovesicles, biomimetic drug carriers, fluorescence correlation spectroscopy, coincident fluorescence burst analysis, atomic force microscopy.

Advances in production of high-affinity antibodies (Ab) harnessed with dedicated pharmacological actions are paving the way for targeting previously untreatable diseases,[1-6] and hold significant potential for the development of novel immunotherapeutic agents boosting the effectiveness of tumor treatments besides chemotherapy.[1, 7, 8] However, most of the available methods for antibody delivery are restricted to extracellular or cell-surface-bound targets,[1-11] and exclude the scope of a growing and important class of antibodies currently being developed against intracellular targets.[8, 10-13] Major difficulties in Ab-deployment against intracellular targets stem from their relatively large size and chemical composition preventing them from naturally crossing the cell membranes and limiting their blood-circulation times and therapeutic action in the absence of appropriate protective encapsulation.[5, 6, 9-14]

Overcoming these challenges is crucial to open up Ab therapies within intracellular spaces. Accordingly, significant research efforts are being devoted to devising efficient methodologies for the delivery of antibodies across immune system and cell membranes, ranging from intracellular injection to camouflaged transport techniques.[14] The former relies on harsh mechanical disruption of the cell membrane through injection or electroporation, with limited loading efficiency and significant impact on cell viability, exclusively suitable for in vitro studies.[15, 16] Alternative approaches involve antibody camouflaging using cell-penetrating peptides, engineered nanoparticles, or liposomes to facilitate antibody transport across cellular membranes.[5, 6, 9-13, 17, 18]. Among these, nanocarrier-assisted delivery, employing polymeric nanoparticles, [5, 19] lipid nanovesicles, [20] and nanoparticles camouflaged with the aid of biomimetic coatings derived from cell membranes [21-25] stand out as promising approaches for drug delivery. Liposomes, known for biocompatibility and controlled release properties, face limitations due to protein corona formation and short-term cargo preservation effects.[26] Some challenges can be mitigated by PEG-polymerization,[19] which however may trigger anti-PEG immunoglobulin production in-vivo, resulting in lowered blood circulation times and degraded immunogenicity.[27, 28]

The innate biocompatibility and non-immunogenicity of red blood cell (RBC) membranes makes them ideal raw materials for direct use as bio-camouflaging materials in a variety of treatments and as drug carriers for intracellular delivery, in nanovesicle forms.[24, 29, 30] RBC membrane-coated nanocarriers have been already studied for Ab delivery and proven to afford longer circulation times thanks to functional RBC-membrane proteins such as CD47.[22, 23, 28, 29, 31] However, the use of such RBC-camouflaged nanocarriers requires the Ab cargo to be aggregated first into a solid form,[21, 25] which may compromise its functionality and induce complications.[32] Such drawbacks can be overcome by drug carriers directly synthesized from RBC-membranes in the form of nanovesicles, provided that suitable procedures become available for their loading with antibodies. [33, 34]

Recently a novel methodology was devised for synthesizing and loading RBC-derived nanovesicles, similar to exosomes, enabling large-scale production in stable formulations with engineerable properties. This technique, initially applied for loading dUTP cargo molecules [35], has now been extended to demonstrate the loading of RBC membrane-derived nanovesicles with larger molecular cargos, such as goat-anti-chicken IgY (H+L) secondary antibodies with significantly larger molecular weights (~145 KDa) than labeled dUTP (~1 KDa). This study systematically analyzes and quantitatively compares the results of Ab-loading with dUTP-loaded vesicles under identical processing conditions, employing spectroscopic protocols developed for single-vesicle profiling with single-molecule resolutions.[33, 35] The findings reveal that Ab-loading yields are maximized for slightly (∼5-10 nm) larger vesicle radii than the ones of dUTP-loading, consistent with the smaller size of the latter, yet still in the ∼50 nm radius range typical of exosome-mimetic nanocarriers. Additional cleaning of nanocarrier solutions using an exosome spin column shows comparable average loading yields of 14% for both Ab and dUTP. The inferred average number of cargo molecules loaded in each nanovesicle features also very similar values (2.25 for Ab and 2.49 for dUTP), exceeding two in both cases, despite their large size discrepancy. The results provide clear evidence for the viability of human erythrocyte-derived nanovesicles for Ab-loading and pave the way to their exploitation as a novel biomimetic system for potential antibody delivery.

Figure 1a provides a flow chart for the preparation of antibody-loaded nanovesicles, closely following previously developed procedures for dUTP cargo molecules.[36] The method involves multiple ultracentrifugation steps to purify RBC ghosts from human blood,[37] and isolate exosome-like vesicles from detergent-resistant membrane (DRM) solutions at a buoyancy of 30% sucrose (1.13 g/cm3) in sucrose gradients (details in S1). The Ab-loading is performed through post-hypertonic lysis of RBC vesicles,[38] inducing vesicle rupture and their subsequent revesiculation, upon which they may engulf Ab-molecules deliberately dispersed in physiologic buffer (see also S1). For this pilot study, a goat-anti-chicken IgY (H+L) antibody is chosen as the cargo molecule. The antibody is conjugated with AlexaFluor®488 (Thermo Fisher), for green fluorescence tagging in optical characterizations using dual-color fluorescence microscopy (DCFM). The outer membranes of the nanovesicles are further stained with CellVue®Claret (Sigma-Aldrich), for far-red fluorescence. The dual-color green and red tagging scheme is illustrated in Fig. 1b. As highlighted in Fig. 1a, the sample preparation process encompassed two slightly different sample typologies of nanovesicles, denoted as RBC and $RBC^+$. The main difference between the two consists in an additional cleaning step performed at the end on the latter ($RBC^+$), with an exosome spin column purification procedure,[39] as detailed in S1. In all cases, loaded and tagged RBC or $RBC^+$ nanovesicles, dispersed in PBS solution, underwent systematic characterizations by means of atomic force microscopy (AFM, Fig. 1c-d) and confocal fluorescence microscopy (Fig. 1e-f), according to experimental and analytical protocols originally defined in previous publications.[35, 40]

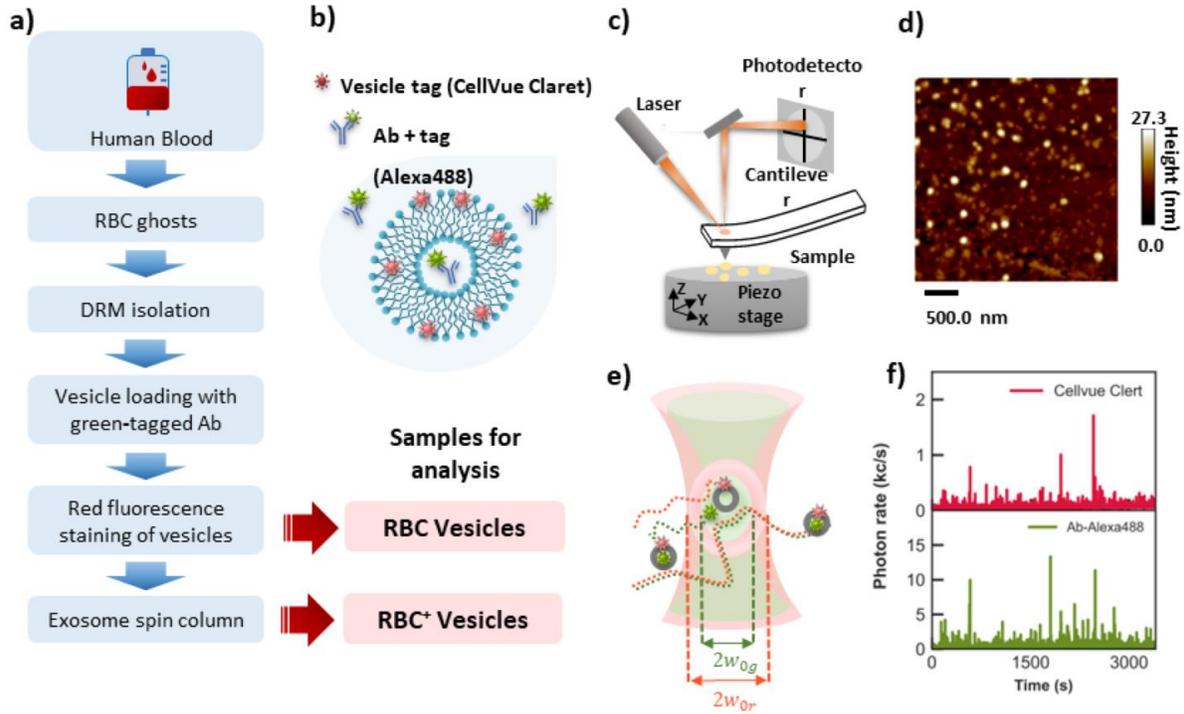

**Figure 1.** a) Preparation steps of antibody loaded nanovesicles. Ab = antibody, RBC=red blood cell nanovesicles, RBC[+]= RBC with additional solution cleaning. b) Dual-color fluorescent staining scheme for the Ab cargo (Alexa488, green dye) and the nanovesicle (CellVue Claret, red dye). Atomic force microscopy (AFM): c) experimental setup and d) image of nanovesicles. Dual-color fluorescence microscopy (DCFM): e) setup and f) typical time traces in detection for the red (vesicles) and the green (antibody molecules) signal channels. Coincident red and green temporal bursts denote Ab-loaded nanovesicles.

The considerable heterogeneity of the exosome-mimetic nanovesicles under investigation necessitates characterizations at the single-vesicle level to extract critical physical parameters, such as size distribution, and facilitate thorough assessments of antibody loading. To address these challenges, we employ experimental methodologies that involve concurrent AFM and DCFM measurements (Fig. 1c-f) and dual-color coincident fluorescence burst (DC-CFB) analyses for size-resolved characterizations of both carrier nanovesicles and their cargos. Figure 2 presents key outcomes of these characterizations conducted on the overall populations of RBC and RBC$^+$ nanovesicles following their synthesis and Ab-loading procedure.

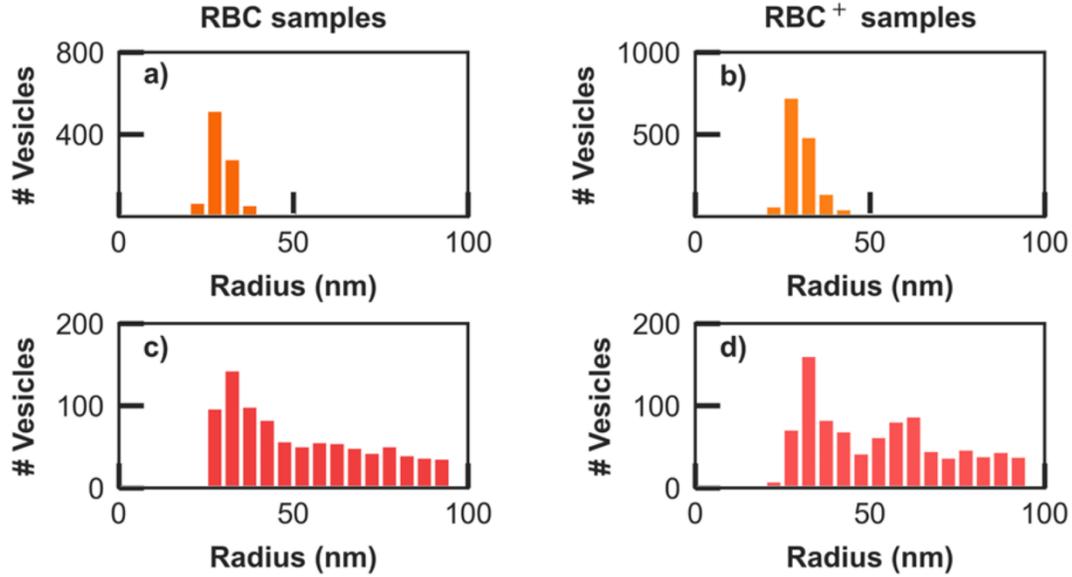

**Figure 2.** Size histograms of the whole populations of nanovesicles assessed through a-b) AFM and c-d) red burst analysis in fluorescence microscopy experiments, for RBC (plots a and c) and RBC$^+$ (plots b and d) sample preparations.

Detailed AFM investigations confirm the formation of single DRM nanovesicles for both sample typologies (RBC and RBC$^+$), featuring size ranges and distribution profiles akin to those of exosomes and exosome-mimetic nanovesicles.[35, 41-43] As illustrated by the plots of Fig. 2a and b, and further quantified by the data in Table I, the AFM histograms for the two sample preparations appear to peak at the same vesicle radius, i.e.: $R_{max}^{AFM} = 27\ nm$, and exhibit comparable values of their average radius, i.e.: $R_{ave}^{AFM} = 30$ nm for RBC, and 31 nm for RBC$^+$ samples. Complementary insights on the hydrodynamic size populations of two nanovesicles are obtained by confocal fluorescence microscopy experiments (SI), through a burst analysis of the red fluorescence signal time traces, yielding the distributions shown in Fig. 2c and 2d, for RBC and RBC$^+$ samples, respectively. For a direct comparison, Table I reports the values retrieved by AFM and fluorescence measurements for both preparations subject to Ab- (rows 1-2) and dUTP- (rows 3-4) loading processes. The latter (see also SI) were processed at the same time and under identical experimental conditions, for a direct comparison with the Ab-loading cases and to provide a reference to previous literature.[35]

|  | AFM | | Fluorescence microscopy | |
|---|---|---|---|---|
| **Sample typology** | $R_{max}$ | $R_{ave} \pm \sigma_R$ | $R_{max}$ | $R_{ave} \pm \sigma_R$ |
| Ab-RBC | 27 nm | 30 ± 4 nm | 32 nm | 59 ± 26 nm |
| Ab-RBC$^+$ | 27 nm | 31 ± 5 nm | 32 nm | 60 ± 26 nm |
| *dUTP*-RBC | *27 nm* | *32 ± 8 nm* | *27 nm* | *49 ± 22 nm* |
| *dUTP*-RBC$^+$ | *27 nm* | *30 ± 4 nm* | *32 nm* | *56 ± 25 nm* |

**Table I.** Summary of the statistics of the RBC and RBC$^+$ nanovesicle populations subject to the loading procedures of Fig. 1a, with Ab or dUTP as cargo molecules, retrieved from AFM and red fluorescence burst analyses. $R_{max}$ = nanovesicle radius at the peak of their size-distribution, $R_{ave}$ = average radius for each vesicle population, $\sigma_R$ = standard deviation of the vesicle radius.

DCFM experiments, as outlined in S1, provide comprehensive insights into nanovesicle populations and their loading.[35] The burst analysis of time traces from the red membrane dye (Fig. 1b) in the fluorescence experiments (Fig. 1c) allows for the retrieval of size-dependent statistics for the entire nanovesicle populations, as depicted in Fig. 2c-d. Fluorescence-derived estimates for size of nanovesicles tend to be slightly larger than those obtained by AFM. This discrepancy is attributed to the larger hydrodynamic size of vesicles in physiological solution, the setting for fluorescence microscopy measurements, compared to dry conditions used for AFM. The observed shift in the peak radius of fluorescence and AFM distributions is relatively small ($R_{max} - R_{max}^{AFM} \sim 5\ nm$). However, the difference becomes more pronounced when considering the average values of vesicle radii ($R_{ave} - R_{ave}^{AFM} \sim 30\ nm$) are likely due to the random diffusion trajectories through the detection volume and the tendency of biomimetic nanovesicles to aggregate in physiological solutions, as consistent with prior reports.[35] This effect, documented in the literature,[44] is confirmed by the longer tails in the distributions obtained from fluorescence data, particularly visible for $R \gg 50\ nm$ in Fig. 2c-d and essentially absent in the narrower AFM profiles of Fig. 2a-b. Furthermore, both AFM and fluorescence results consistently indicate no significant impact of the additional cleaning step (RBC vs RBC$^+$) on Ab-loaded sample preparations. The maximally populated nanovesicle radius derived from the fluorescence data analysis remains the same for both RBC and RBC$^+$ samples ($R_{max} = 32\ nm$) and this is equally true for their average radii ($R_{ave} \sim 60\ nm$).

A notable observation from comparing Ab and dUTP loading results is the larger vesicle size associated with antibody loading. This aligns with the substantial weight difference between Ab molecules and dUTP, with the former being over two orders of magnitude higher-molecular weight than the latter. A distinct difference is observed in the values of $R_{max}$ and $R_{ave}$ after applying the extra cleaning procedure (RBC vs RBC$^+$) to dUTP-loaded samples, an effect not observed in antibody loading. In the dUTP case, both $R_{max}$ and $R_{ave}$ show an increase of approximately 5-7 nm post-cleaning. This suggests additional size-filtering effects during the exosome spin column process, potentially favoring slightly larger vesicles and better matching the size-distribution peak ($R_{max}$) of Ab-loaded samples, which is approximately 5 nm larger than in the dUTP case. The impact of cleaning (RBC vs RBC$^+$) is more pronounced in the case of dUTP loading compared to Ab loading, significantly affecting also the retrieved loading yields, as discussed in the next section.

Figure 3 illustrates the result of further investigations into the sub-populations of loaded nanovesicles performed by dual-color coincident fluorescence burst (DC-CFB) analysis, considering nanovesicles loaded with antibodies and their dUTP-loaded counterparts (SI and Ref.

35). Fig. 3a (b) shows the size distribution of Ab-loaded RBC (RBC$^+$) nanovesicles, while Fig. 3c (d) illustrates their loading yield, i.e.: $\eta(R) = \frac{N_{load}(R)}{N_{tot}(R)}$, quantified as the ratio of the number of loaded nanovesicles ($N_{load}$), determined from coincident green and red bursts, (Fig. 3a-b) and the total nanovesicle count ($N_{tot}$), determined from red burst analyses (Fig. 2c-d), as a function of the nanovesicle radius $R$. Table II summarizes key figures of merit extracted by the DC-CFB analysis of the experiments, to enable quantitative comparisons between RBC and RBC$^+$ preparations, as well as loading with Ab and dUTP cargo molecules.

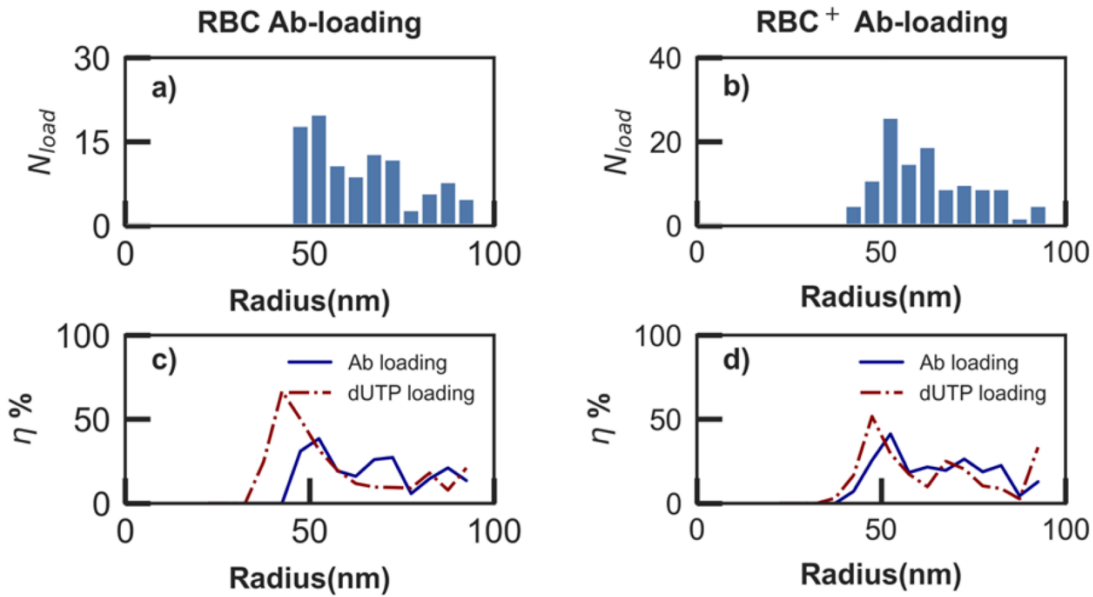

**Figure 3.** Histogram of the number of Ab-loaded nanovesicles ($N_{load}$) as a function of nanovesicle radius $R$, for: a) RBC and b) RBC$^+$ sample preparations. Size-distribution of the loading yield $\eta(R)$, for Ab (solid lines) and dUTP (dashed lines) cargo molecules, for : c) RBC and d) RBC$^+$ preparations.

|  | $R_{max}^{load}$ | $R_{ave}^{load} \pm \sigma_R$ | $\eta_{max}$ | $\eta_{av}$ | $N_{max}$ | $N_{ave} \pm \sigma_N$ |
|---|---|---|---|---|---|---|
| **Ab-RBC** | 52 nm | 66 ± 15 nm | 38 % | 14 % | 1.75 | 2.25 ± 0.75 |
| **Ab-RBC$^+$** | 52 nm | 65 ± 15 nm | 41 % | 14 % | 2.25 | 2.53 ± 0.76 |
| *dUTP-RBC* | 42 nm | 55 ± 17 nm | 67 % | 20 % | 2.25 | 2.71 ± 0.71 |
| *dUTP-RBC$^+$* | 48 nm | 61 ± 17 nm | 52 % | 15 % | 2.25 | 2.49 ± 0.54 |

**Table II.** Summary of the statistics of the loaded sub-populations of RBC and RBC$^+$ nanovesicles, with Ab and dUTP cargo molecules, retrieved from dual-color coincident fluorescence burst analyses. $R_{max}^{load}$ = loaded nanovesicle radius at the peak of their size distribution, $R_{ave}^{load}$ = average radius of the loaded vesicles, $\sigma_R$ = standard deviation of the vesicle radius. $\eta_{max}$ = loading yield for $R=R_{max}^{load}$, $\eta_{ave}$ = average loading yield. $N_{max}$ and $N_{ave}$ are the maximum and average number of loaded cargos per vesicle ($N_{Ab}$ or $N_{dUTP}$), respectively.

Similar to dUTP-loaded nanovesicles, the Ab-loaded nanovesicle populations exhibit a skewed distribution in radius, with a primary peak at $R_{max}^{load}$< 70 nm and an extended tail toward larger sizes (>100 nm), where experimental artifacts due to vesicle agglomeration combined with random diffusion trajectories become prominent. The size distributions of Ab-loaded nanovesicles in RBC (Fig. 3a) and RBC+ (Fig. 3b) samples show essentially the same values for the peak ($R_{max}^{load}$=52 nm) and average ($R_{ave}^{load}$~65 nm) radii, indicating negligible impact of the solution cleaning step. Comparing the histograms in Fig. 3a-b with those of Fig. 2c-d, depicting the total nanovesicle populations for the Ab-loading case, highlights a shift of the vesicle distributions toward larger sizes upon loading. This shift is quantified by comparing the values for $R_{max}$ in Table I and $R_{max}^{load}$ in Table II, revealing an increase of ~20 nm in the peak radius of Ab-loaded compared to the whole nanovesicle populations. This size increase in the loaded nanovesicle population is also apparent in the values of the average radii of Ab-loaded ($R_{ave}^{load}$, Table II) and whole ($R_{ave}$, Table I) vesicle populations. In comparison to dUTP-loaded vesicles, the Ab-loaded vesicles exhibit approximately 10 nm-larger average and peak sizes, consistent with the larger size and molecular weight (~145 KDa) of antibodies compared to labeled dUTP molecules (~1 KDa). Another notable difference is observed in the dUTP-loaded vesicle distributions after the extra-cleaning process, evident from the data in S3 and Table II, indicating an increase by 6 nm in both $R_{max}^{load}$ and $R_{ave}^{load}$ of RBC+ versus RBC dUTP-loaded nanovesicles. This consistent shift to larger sizes also occurs in the statistics of the overall nanovesicle populations when comparing RBC and RBC+ preparations in the dUTP case, highlighting the size-filtration effect of the original nanovesicle populations associated with the RBC+ cleaning step, favoring slightly larger vesicles (R≥50 nm) that better match Ab-loaded vesicles. This justifies the observed changes affecting the dUTP-loaded and not the Ab-loaded vesicles, as well as the modification of the dUTP-loading yield results in RBC and RBC$^+$ preparations, apparent in Fig. 3c-d (dashed lines) and in Table II ($\eta_{max}$ and $\eta_{ave}$ in rows 3-4).

Further analysis of the dual-color experimental data, explained in SI, enables size-resolved evaluations of loading yields, as depicted in Fig. 3c and 3d for RBC and RBC$^+$ samples, respectively. Consistent with previous discussions, the Ab-loading yield distribution, $\eta(R)$, remains unaffected by the cleaning procedures, peaking at the same vesicle radius, $R_{max}^{load} = 52\ nm$, for both RBC and RBC$^+$ preparations. The maximum loading yield, $\eta_{max} = \eta(R_{max}^{load})$, is also minimally affected, with values of 38% and 41% for RBC and RBC$^+$ samples, respectively (Table II). Average values of loading efficiencies and vesicle sizes exhibit similar trends, with $\eta_{ave} = 14\%$ and $R_{ave}^{load} \sim 65\ nm$, respectively, regardless of the extra cleaning step in the Ab case. However, this is not observed for the dUTP case, as evident in the loading yield distributions for RBC and RBC$^+$ preparations (dashed lines in Fig. 3c and 3d) and the corresponding figures of merit in Table II. The RBC$^+$ cleaning step induces a clear modification of the peak yield, with $R_{max}^{load}$ increasing from 42 to 48 nm and $\eta_{max}^{load}$ decreasing from 67% to 52%, along with substantial effects on average values, with $R_{ave}^{load}$ increasing from 55 to 65 nm and $\eta_{ave}$ decreasing from 20% to 15%. These trends align with those highlighted in the vesicle populations for dUTP cargo molecules, indicating a more pronounced influence of the additional cleaning process and its associated size-filtering effect, as discussed with reference to Fig. 2 and Table II. This suggests promising avenues for optimizing sample preparation to maximize loading efficiency in drug delivery applications, involving size-based filtering of nanovesicle populations affected by cargo molecule loading, as further discussed in the following section.

Finally, the single-molecule resolving capability of fluorescence measurements, combined with further analyses and calibration experiments detailed in S2 and Ref. 35, afforded also statistical investigations on the number-normalized brightness per vesicle ($N_{Ab}$), illustrated in Fig. 4a-b. Equivalent data for the dUTP case are presented in S3 (Fig. S11), and a comparative summary of the results is provided by Table II (columns 6 and 7), listing the retrieved values of the maximum number of cargos per loaded nanovesicle ($N_{max}$) and its average ($N_{ave}$) for all four sample typologies.

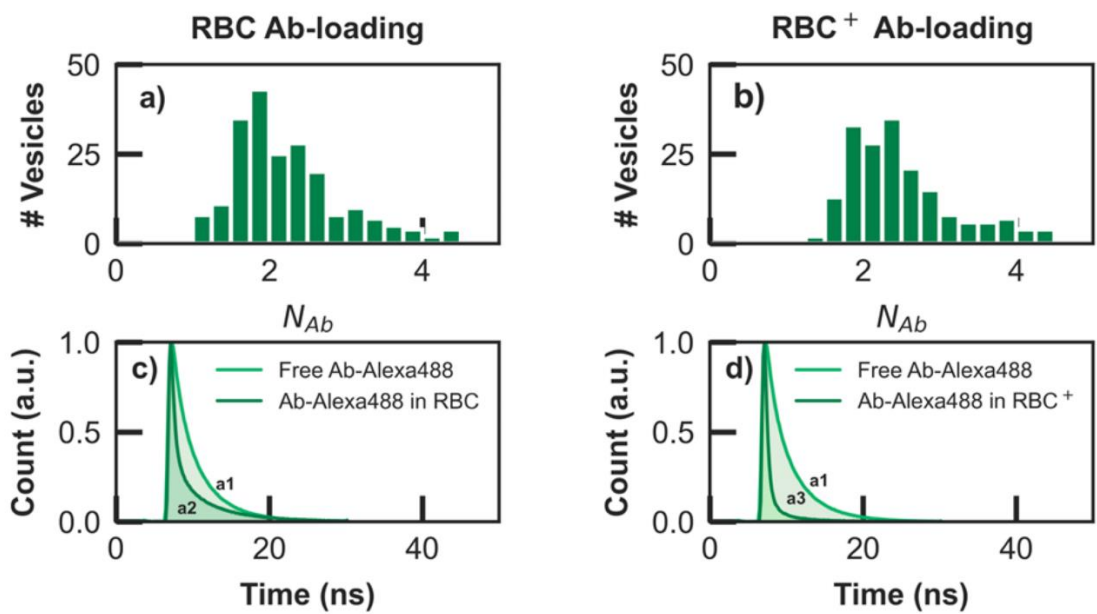

**Figure 4.** Histograms of a number-normalized brightness per vesicle ($N_{Ab}$), for : a) RBC and b) RBC$^+$ sample preparations. The normalized lifetime histograms of Alexa488 dye bound to Ab-cargo molecules in three conditions such as a$_1$: free in solution, a$_2$: loaded into c) RBC nanovesicles, and a$_3$: loaded into d) RBC$^+$ nanovesicles.

The analysis shows that regardless of the type of cargo (Ab or dUTP) and solution cleaning procedure (RBC or RBC$^+$), the loaded nanovesicles contain on average two cargo molecules, with excellent agreement between the values obtained for Ab and dUTP in RBC$^+$ samples, namely: $N_{max} = 2.25$ and $N_{ave} = 2.5$, confirming the consistency and reliability of the antibody loading process into red-blood-cell-derived nanovesicles.

Moreover, the calibration measurements required for the retrieval of the statistics of cargo molecules per nanovesicles, as described in S2, revealed unexpected features in the fluorescence signals of Alexa488 dye bound to cargo molecules entrapped into the nanovesicles. Fig. 4c-d (Fig. S11c-d) illustrates these findings, displaying lifetime histograms of green fluorescence signals from Alexa488 dye bound to Ab (dUTP) molecules in three conditions: 1) free in solution, 2) loaded into RBC nanovesicles, and 3) loaded into RBC$^+$ nanovesicles. The results clearly show a reduction in fluorescence lifetimes for Alexa488 when it is encapsulated in the nanovesicles, whether bound to Ab or dUTP. The observed shortening of lifetimes in Fig. 4c-d, which is considered in the results derivation, is a noteworthy effect not highlighted in previous studies so far. Unlike free molecules with monoexponentially decay, $\tau_{lifetime} = 4$ ns for Alexa488 with dUTP and 3.6 ns with Ab, entrapped molecules exhibit a continuous range of shorter lifetimes possibly due to potential FRET interactions with hemoglobin inside RBC nanovesicles. Additional centrifugation steps during exosome spin column cleaning, further exacerbate this effect, resulting in the lowest lifetimes of fluorescent tags particularly in the RBC$^+$ case. The number and interaction distance of hemoglobin molecules are unpredictable and uncontrollable, leading to varying lifetimes. To account for this effect on the fluorescence brightness of the tagging dye inside RBC and RBC$^+$ vesicles (see also S2), the area under their normalized lifetime histograms ($a_2$ and $a_3$ in Fig. 4c and d) is compared with that of free cargo molecules ($a_1$ in Fig. 4c-d), allowing for the calibration of the average photon count for the entrapped cargos (see Table S3). Conclusively, the DC-CFB assessments provide the two-dimensional profile of loaded vesicles versus radius and Ab- number (number-normalized brightness) per loaded vesicle, as illustrated in Figure 5, revealing the mostly populated sizes and the load extent for the Ab-loaded nanovesicles.

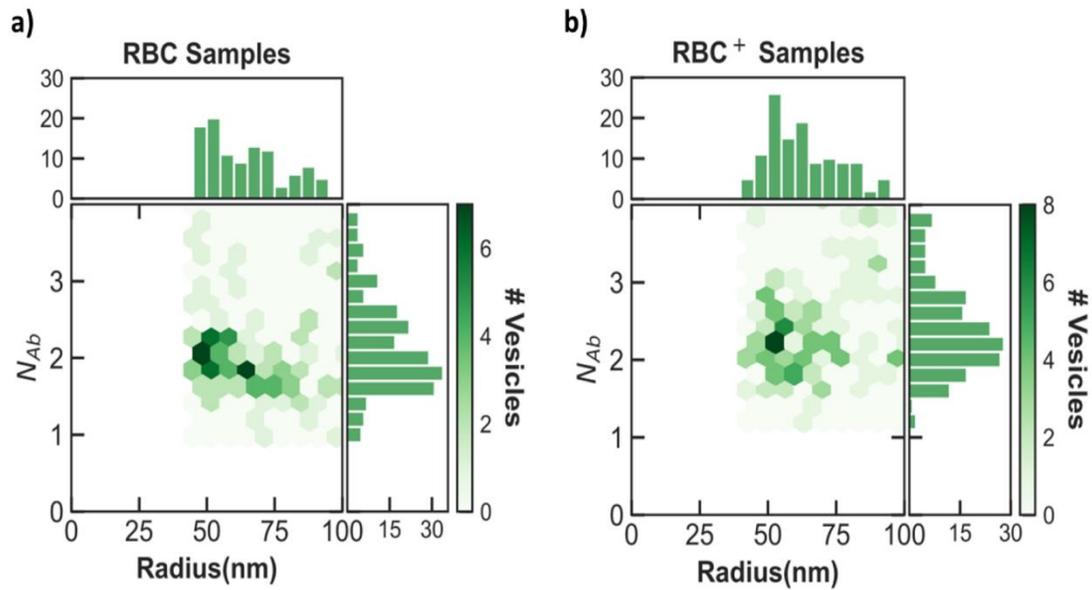

**Figure 5.** Profile of antibody-loaded nanocarriers: Two dimensional histograms of loaded nanovesicles versus their radius ($R$) and number-normalized brightness per vesicle ($N_{Ab}$), obtained from dual-color coincident fluorescence burst assessments. The size and number of Ab-cargo molecules per vesicle for the loaded vesicles are depicted in the horizontal and vertical subplots, respectively, for a) RBC and b)RBC[+] preparations. It demonstrates that loaded nanovesicles are mostly populating at radius of 52 nm and maximum brightness-normalized number of 1.75 and 2.25 Abs per RBC and RBC[+] nanovesicles, respectively.

In summary, we have demonstrated successful antibody-loading into synthesized nanovesicles from red-blood-cell membranes. Dual-color fluorescence burst investigations at single-vesicle level revealed a preference for nanovesicles with an average radius of around 65 nm when loaded with Ab-cargos, compared to 55 nm obtained in reference experiments performed on dUTP cargos, which aligns with the larger size and molecular weight of Ab molecules. Loading yields for Ab-cargos were comparable to those of dUTP-loaded vesicles, peaking at approximately 40%. The optimal vesicle radius of 52 nm and an average loading efficiency of 14% was obtained for Ab-loaded nanovesicles. Considering unexpected lifetime shortening of Alexa488 fluorophore in produced nanovesicles, likely due to FRET interactions with hemoglobin molecules, revealed an average loading of 2.25 antibody molecules per nanovesicle, consistent with dUTP cargo results under identical conditions. Additional cleaning procedures also provided stable results for Ab-loading and insights into the relationship between nanovesicle physical properties and cargo nature, offering opportunities to enhance production yields and customize these nanovesicles for specific therapeutic Ab-agent delivery.

**ASSOCIATED CONTENT**

**Supporting Information**. Supporting information is available free of charge at http://pubs.acs.org and includes details on the experimental methods, data analysis procedure and summary of additional reference results.

AUTHOR INFORMATION

**Corresponding Author**

Maryam Sanaee, *email: msanaee@kth.se, ORCID: 0000-0001-8661-6583.

**Author Contributions**

K. G. and M. S. conceived the study. K. G. R. was responsible for the parts of the study that involved identity, biology, design and loading of the nanovesicles from two different sources. K. G. R. and J. M. M. prepared the samples. M. S. performed the AFM measurements and their analysis. J. W. and E. S. developed the DCFM setup. M. S. and E. S. performed the dual-color fluorescence measurements and M. S. developed and performed the dual-color CFB coding and data analysis. M. S. and K. G. wrote the manuscript with contributions and approval from all authors.

**Funding Sources**

The work was supported by the program for biological pharmaceutics of the Swedish Innovation Agency (Vinnova grant no 2017-02999), and by the OQS Research Environment for Optical Quantum Sensing of the Swedish Research Council (VR grant no 2016-06122). K. G. gratefully acknowledges further support from the Knut and Alice Wallenberg Foundation through the Wallenberg Center for Quantum Technology (WACQT) and from the Swedish Research Council via grant VR 2018-04487. J. W. acknowledges further support from the Swedish Research Council via grant VR 2021-04556.

## ACKNOWLEDGMENT

We thank the blood-donors organization in Uppsala for the blood samples.

**Declaration of interest statement.**

The authors report no competing interest.

## ABBREVIATIONS

Ab, Antibody; AFM, atomic force microscopy; CFB, Coincident Fluorescence Burst; DCFM, dual-color fluorescence microscopy; DRM, Detergent Resistant Membranes; RBC, red blood cell.

# Supplementary Information

# Antibody-loading of biological nanocarrier vesicles derived from red-blood-cell membranes

*Maryam Sanaee,[1]\* K. Göran Ronquist,[2] Elin Sandberg,[1] Jane M. Morrell,[2] Jerker Widengren,[1] and Katia Gallo[1]*

[1] Department of Applied Physics, School of Engineering Sciences, KTH Royal Institute of Technology, Stockholm 10691, Sweden.

[2] Department of Clinical Sciences, Swedish University of Agricultural Sciences, Uppsala 75007, Sweden

*Corresponding author. Email: msanaee@kth.se

# Contents

# S1- Experimental methods

A summary of the experimental methods employed during this research are provided in following subsections including the preparation description of the sample preparation, the AFM and dual-color fluorescence microscopy measurements.

## S1-1- Sample preparation

The preparation of Ab-loaded red blood cell (RBC) nanovesicles includes five main steps and is briefly described below. They follow closely the previously developed procedures with dUTP-cargo molecules [1].

### a) Red Blood Cell ghosts' preparation

Following the acquisition of blood bags containing red blood cells from Uppsala University Hospital, a 10 mL volume of red blood cells was washed 3 times (1:5 ratio) in phosphate buffered saline (PBS), by centrifugation (Nino lab Heraeus) at 2100g and 4°C for 10 min. Washed red blood cells were lysed in hypotonic phosphate buffer (53.4 mOsmol.L$^{-1}$). Most of the hemoglobin, was removed by cyclic steps of centrifugation/dilution at 20,000g and 4°C for 30min, Beckman Coulter (BC), SW32Ti rotor) Derived RBC ghosts were stored at -20°C and utilized within a timeframe of less than three months.

### b) DRM vesicles isolation

Stored RBC ghosts (10 mL) were pelleted by ultracentrifugation at 100,000g and 4°C for 1h, (BC, SW32Ti). The pellet was dissolved in PBS containing 1% triton x-100 and incubated on ice for 30 min before separation on density gradient built by 40%, 30%, 24%, and 10% sucrose, run at 230,000g and 4°C for 5h, (BC, SW40Ti rotor). Fractions on top of 24% and 30% sucrose (density of 1.10 and 1.13 g.cm$^{-3}$) were collected and in later separations the 24% density was skipped. Extracted fractions were pelleted in PBS containing 10% sucrose at 150,000g and 4°C for 1h, (BC, SW32Ti). Pellets, sucrose charged detergent resistant membrane nanovesicles, DRM-nanovesicles, were stored at -20°C.

### c) DRM-nanovesicle loading with Alexa 488 antibody

The central procedures involved in loading Ab cargo into DRM-nanovesicles, following the final preparation steps are illustrated in Figure S1. The frozen DRM-nanovesicle pellet was dissolved in 10µL goat anti-Chicken IgY (H+L) secondary antibody tagged with Alexa488 (Ab-Alexa488 ®ThermoFisher), and after a short incubation diluted in PBS. The shift from hypertonicity caused by encapsulated sucrose to isotonicity due to Ab:PBS dilution led to "post-hypertonic lysis" [2], triggering the rupture of nanovesicles, followed by re-vesicultion in the Ab enriched PBS solution, all fluorescently labeled samples were shielded from direct light exposure. Ab-loaded DRM-nanovesicles were pelleted in low sucrose (1-3%) by ultracentrifugation at 100,000g and 4°C for 1h (BC, SW32Ti) to remove excess fluorescence.

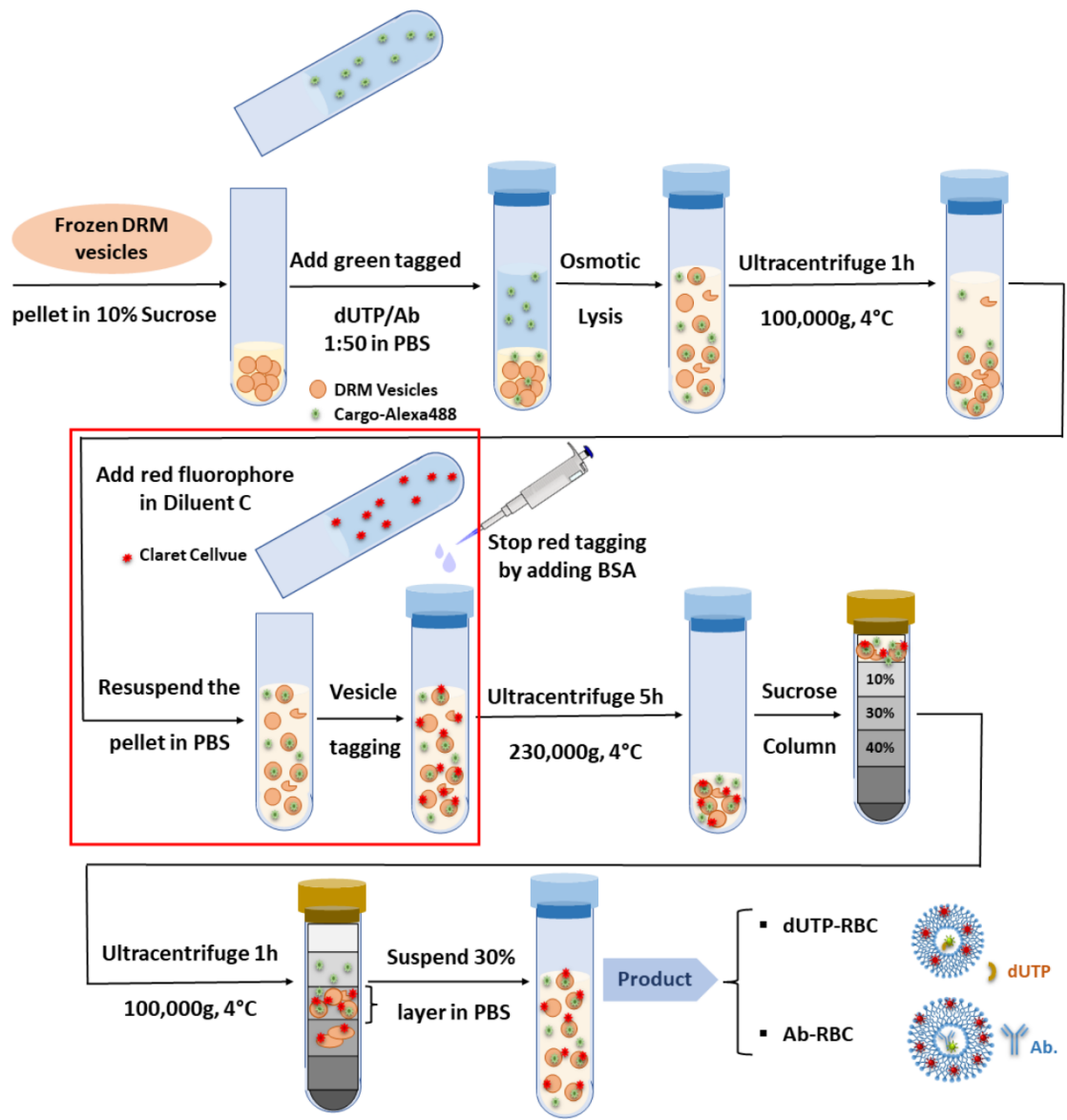

**Figure S1.** The schematic of the core processes associated with the loading of Ab-Alexa488 cargos into DRM nanovesicles derived from RBC-membrane, including the final preparation stages such as red fluorescence tagging of vesicles, highlighted in red box.

**d) Red fluorescence staining of** DRM-nanovesicles

In Figure S1, the process for staining loaded nanovesicles with red fluorescence is highlighted within a red box. To label the vesicle membranes, the pellets were mixed with CellVue Claret membrane kit-staining-component Diluent C (Sigma-Aldrich, Miniclaret-1kt), following the manufacturer's instructions. The staining reaction was stopped by 3% bovine serum albumin in PBS and immediately pelleted by top-loading on a low concentrated sucrose solution (1-3%) at 100,000g and 4°C for 1h (BC, SW32Ti). The pellets of DRM-nanovesicles preparation were resuspended in PBS and kept at 4°C in dark for physics experiments to be conducted. Short summary, this approach is applied to investigate the efficiency of Ab loading of the prepared samples and compare them with dUTP cargo. To reach this goal, the cargo molecules were fluorescently stained by the dye Alexa488 with excitation and detection wavelengths at 485 nm and 535 nm, respectively, while nanovesicle membranes were tagged by the far red dye, CellVue Claret, with corresponding excitation and detection wavelengths of 640 nm and 720 nm, respectively.

**e) Extra cleaning with exosome spin column**

In the last phase of the preparation, to evaluate the impact of additional purification on the loading efficiency and cleanliness of loaded RBC nanovesicles, a portion of these loaded DRM-nanovesicles underwent a purification process utilizing the exosome spin column (ESC, ®Thermo Fisher, ESC MW 3000).

In Figure S2, the real pictures of the prepared samples in different steps are illustrated, including a) the top-loading of purified erythrocyte ghosts in 1% triton x-100 on density gradient and several formed sucrose buoyancies (10%, 24%, 30% and 40%), b) the separated erythrocyte ghosts on density gradient after being treated with 1% triton x-100, while the upper small band is detergent resistant membranes and the broader lower band is bulk erythrocyte membrane which are not floating up as the lighter DRM-fraction, c) DRM-nanovesicles loaded with Ab-Alex488, and finally d) the DRM-nanovesicles stained with far red namely as Ab-RBC.

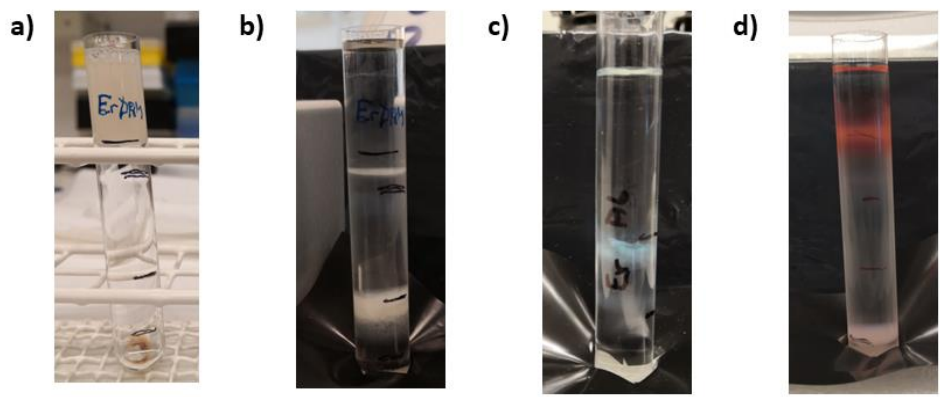

**Figure S2.** The real picture of the samples during each preparation steps encompassing a) the top-loading of purified erythrocyte ghosts in 1% triton x-100 on density gradient, b) the separated erythrocyte ghosts on density gradient after being treated with 1% triton x-100, while the upper small band is detergent resistant membranes and the broader lower band is bulk erythrocyte membrane that are not floating up as the lighter DRM-fraction, c)

DRM-nanovesicles loaded with Ab-Alex488, and d) the DRM-nanovesicles stained with far red as final product (Ab-RBC).

**S1-2- Atomic force microscopy measurements**

The morphology of individual vesicles and their size distribution were assessed using a commercial atomic force microscope (FastScan Bruker AFM). The AFM operated at a slow scan rate of 0.5 Hz and was set to tapping mode under ambient air conditions. NCHV-A (Bruker) cantilevers with a tip featuring an average radius of 8 nm were employed for these measurements. Sample handling and AFM measurements for Ab-loaded samples followed a methodology similar to that recently reported for dUTP-loaded nanovesicles [1]. The AFM-derived size histograms were compared to size histograms of the complete vesicle populations determined through single-color red fluorescence analysis which is explained in next section.

**S1-3- Dual-color fluorescence microscopy experiments**

The dual-color fluorescence microscopy (DCFM) experiments were conducted using a custom-built setup based on a commercial confocal microscope (Olympus FV1200) equipped with a water immersion objective (60x, NA 1.2, Olympus, UPlanSApo) [1]. Figure S2 depicts the main components of this microscope. Two lasers operating at wavelengths of 485 nm (Picoquant LDH-D-C-485) and 640 nm (LDH-D-C-640) were Picosecond-pulsed modulated at 20 MHz repetition rate using pulsed-interleaved mode for sample excitation to avoid dual-color cross talk (Fig. S2 b). The lasers power was adjusted to maintain a consistent excitation of 80 $\mu$W at the back-focal plane of the objective during all measurements. In the detection pathway, the signals emitted by the red and green fluorophores were passed through a 50 µm pinhole and spatially separated by a dichroic mirror. The emitted red photons were further filtered through an HQ720/150 (Chroma) filter, and the green photons passed through an HQ535/70 (Chroma) filter. The fluorescent light emitted at each wavelength was directed onto either Picoquant tau-spad or Perkin & Elmer (SPCM-AQR-14) single photon-counting avalanche detectors (SPADs), connected to a time-correlated single-photon counting (TCSPC) module (HydraHarp 400) equipped with suitable software (Symphotime, Picoquant) for data acquisition and analysis. Prior to each experimental session, calibration measurements were conducted using well-characterized fluorophores like Cy5 and Rhodamine110 as references to confirm the consistency of the measurement settings and to determine the focal detection volume at both red and green excitation wavelengths, which were found to be 0.320 and 0.165 femtoliters, respectively. A volume of 100 $\mu$L (approximately 2 $\mu$g) of the nanovesicle solution (diluted 1:100 in PBS) was applied to an eight-well container on top of the microscope objective. These measurements were carried out for a time duration longer than 50 min at a constant temperature in a completely dark room [1].

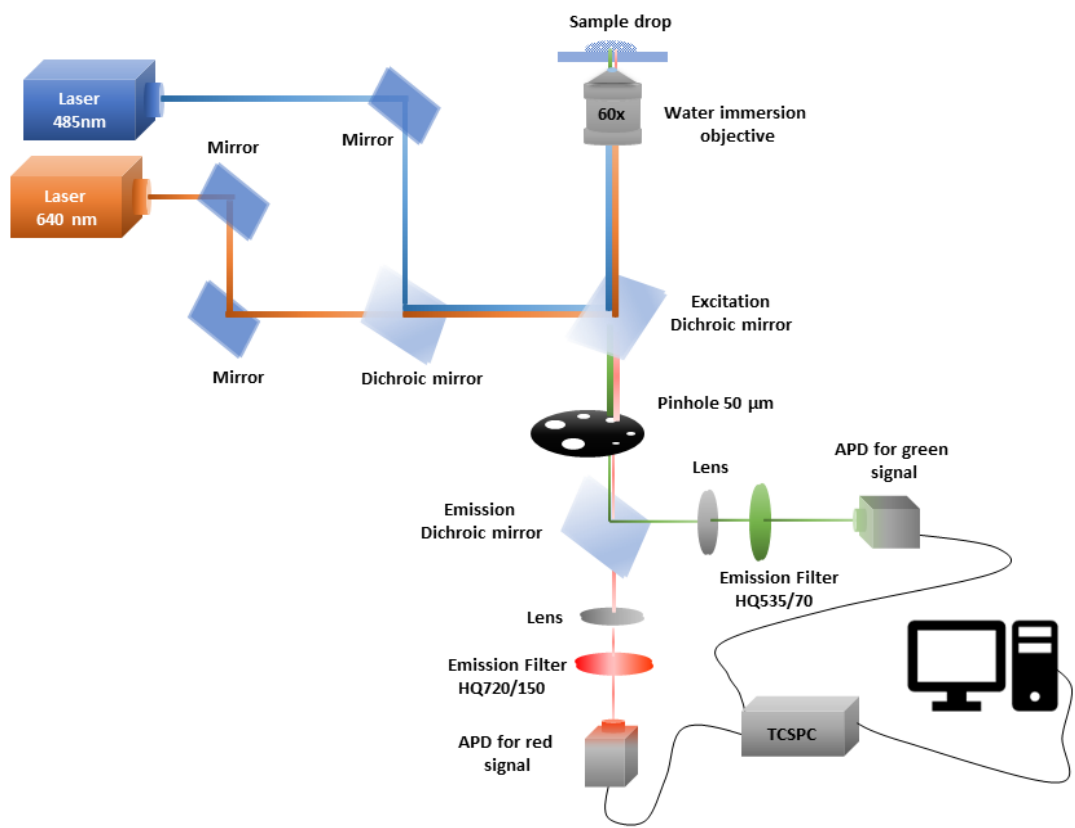

**Figure S3.** a) The essential elements of dual-color fluorescence microscope including two lasers emitting at wavelengths of 485 nm and 640 nm utilizing, the Picoquant LDH-D-C-485 and LDH-D-C-640 lasers, respectively. b) Two lasers were synchronized and operated in a picosecond pulse (shown with sharp red and blue gaussian pulses) modulated at repetition frequency of 20 MHz, utilizing pulsed-interleaved mode, to excite the red and green fluorophores separately in time, while the collected fluorescence signals were time gated through highlighted red and green time windows.

## S2- Data analysis

The dual-color coincident fluorescence burst (DC-CFB) analysis was developed recently [1] and is applied here to rigorously assess the loading efficiency of Ab-RBC and Ab-RBC$^+$ preparations. To that aim, the collected red and green fluorescence signals were time-gated according to their excitation pulses and fluorescence lifetimes, as shown in Figure S4a-d, where the red and green regions highlight the time windows for photon collection in the relevant spectral ranges.

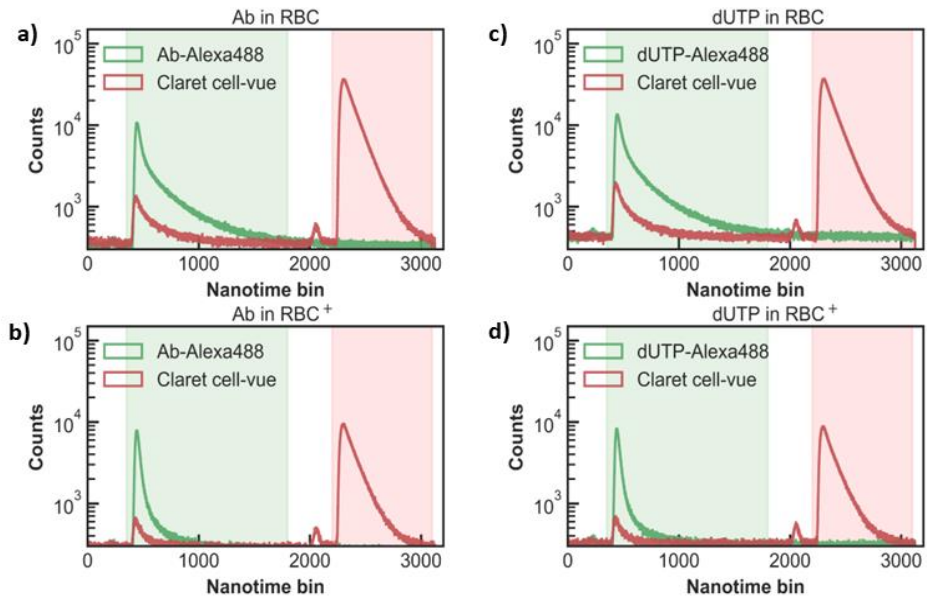

**Figure S4.** The time resolved lifetime histograms of fluorescent green and red channels highlighting their respective lifetimes (Nanotime bin=16 ps) and time gating windows (shaded green and red areas) for: a-b) Ab-loaded and c- d) dUTP-loaded RBC and RBC[+] preparations.

**S2-1- Background calculation**

Before searching for the fluorescent bursts at each color, the background rates versus time were evaluated for each preparation at 50 s temporal intervals, as depicted in Figure S5.

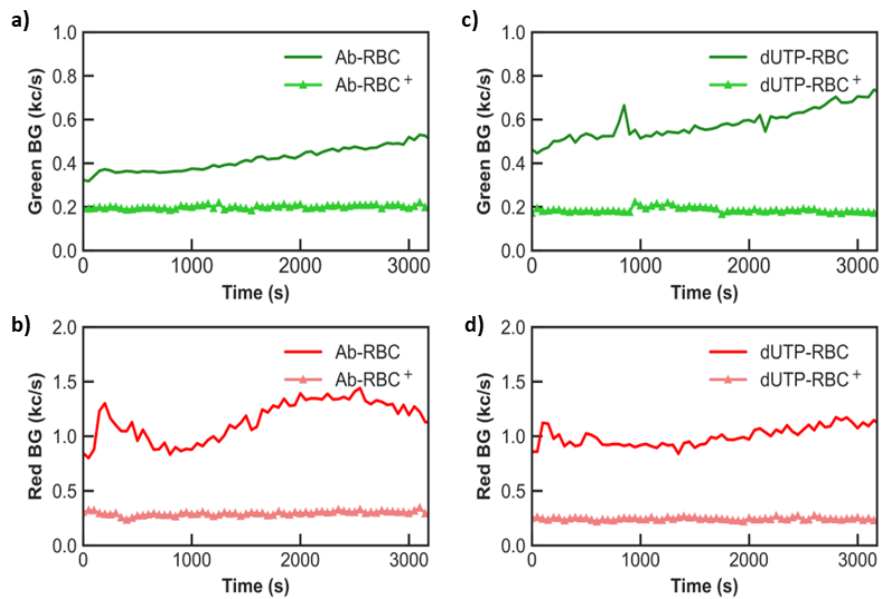

**Figure S5.** Calibration measurements of the background rates (BG) for the green and red channels performed in four different experimental sessions, on: a-b) Ab- and c-d) dUTP- loaded nanovesicles for samples without (RBC, solid lines) and with (RBC[+], line with markers) additional cleaning steps (see S1).

In summary, the evaluations reveal that the average background rates in antibody (dUTP)-loaded vesicles decreased approximately 4.6 (4.4) times in the red channel and 1.95 (2.75) times in the green channel after the additional cleaning process. In conclusion, these results underscore the efficacy of the supplementary cleaning step in eliminating more non-encapsulated cargoes from the final solution. Remarkably, this cleaning process had a substantial impact on the green channel, leading to a fourfold reduction in its corresponding background rates. This reduction is primarily attributed to the removal of non-encapsulated antibody/dUTP cargoes.

**S2-2- Assessment of total population of nanovesicles**

Following the subtraction of time-dependent background rates from the red and green channels, the DC-CFB analytical method was employed to identify the red and green fluorescent bursts, corresponding to the entire nanovesicle population and the green-tagged cargo molecules, whether located inside or outside the RBC nanovesicles, respectively. Based on the Stokes–Einstein diffusion theory, each burst duration time ($\tau_{burst}$) in the fluorescence time traces obtained from the experiments was converted into the radius ($R$) of the nanovesicles or molecules on single-vesicle and single-molecule basis. This conversion was accomplished as $R = \frac{4k_B T}{6\pi\mu w_0^2}\tau_{burst}$, where $k_B$, $T$ and $\mu$ represent the Boltzmann constant, the lab temperature, and the viscosity of the solvent (water), respectively. The lateral radius of the confocal volume, denoted as $w_0$, was measured in calibration experiments as $w_{0r} = 319\ nm$ and $w_{0g} = 254\ nm$ for the red and green laser spots, respectively.

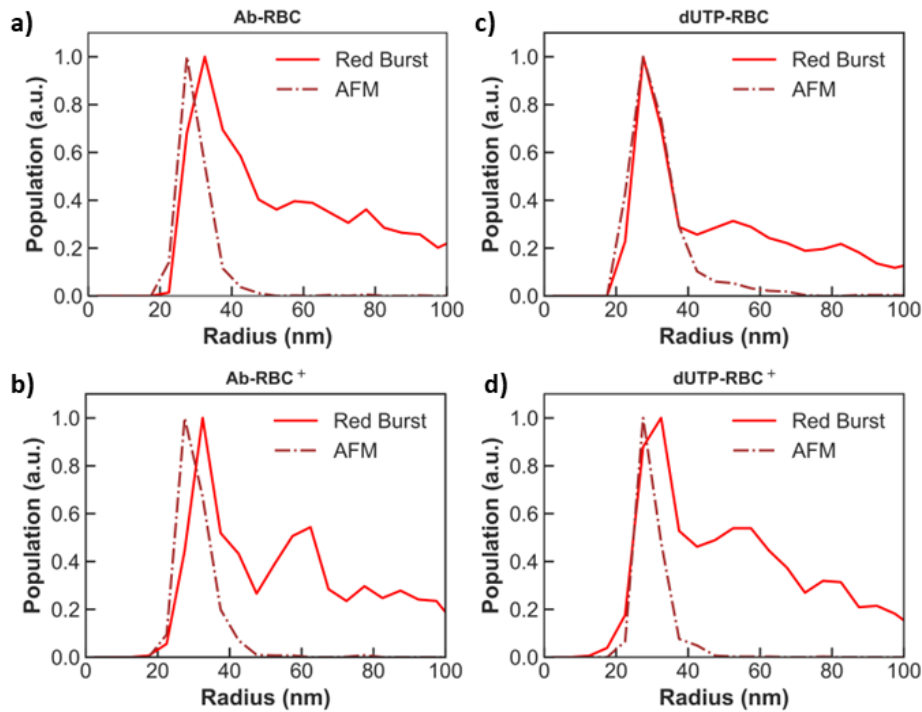

**Figure S6.** Nanovesicle size distributions normalized to their peak value obtained by AFM (dash-dotted brown) and fluorescence red burst analyses (solid red) for: Ab-loaded a) RBC, b) RBC$^+$, dUTP-loaded c) RBC and d) RBC$^+$ preparations.

Therefore, the determination of the total nanovesicle population involved identifying optimized key parameters for the DC-CFB analysis, including minimum photon number (M) and count rate (F) thresholds for both red and green bursts. As previously detailed, the optimal settings for $M_r$ and $F_r$ for the red bursts were established through a burst analysis protocol that systematically compared the red fluorescence results with independent AFM measurements to validate the extracted size distributions. Meanwhile, there might be some differences between AFM and red fluorescence size distributions due to hydration effect in the latter one as it is conducted in physiological buffer solution. The outcomes of the total population assessment using the red fluorescence signal ($N_{tot}$) and their comparison with corresponding AFM size profiles are depicted in Figure S6, for two different preparations (RBC and RBC$^+$) and cargo molecules (Ab and dUTP).

## 2-3- Loaded nanovesicles

The optimized burst conditions for green channel ($M_g$ and $F_g$) were found based on the established protocol in DC-CFB analysis, by searching for coincident-red and coincident-green bursts in synchronized dual-color fluorescence time traces. According to this methodology, the coincident-red and coincident-green size distributions versus vesicles' radius ($R$) were optimized to get minimum difference ($\varepsilon(R)$) between their normalized distributions as their results are depicted in Figure S7 and S7, for both Ab- and dUTP-loaded preparations, respectively. Subsequently, the size distribution of coincident-red is attributed to the sub-population of loaded nanovesicles ($N_{load}$). The summary of the optimized burst conditions for red and green channels are provided in Table S1.

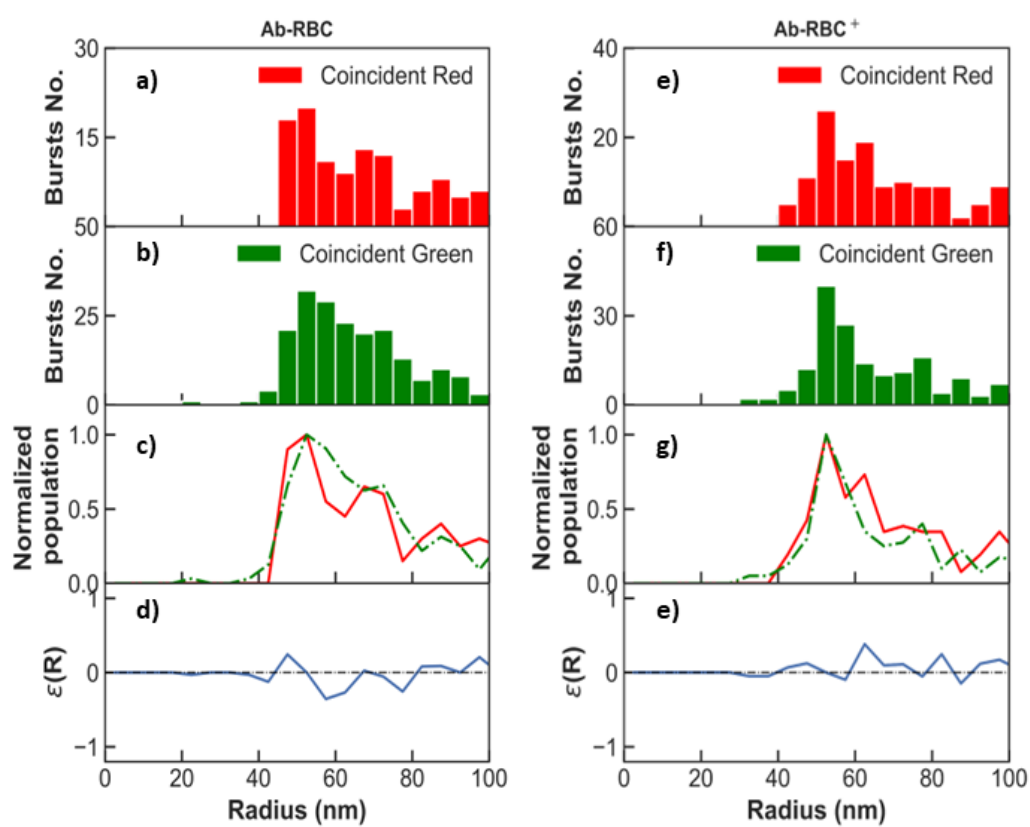

**Figure S7.** Loaded nanovesicle size histograms retrieved from the fluorescence experiments for a-c) Ab-RBC and e-g) Ab-RBC⁺ preparations. a, e) coincident-red and b, f) coincident-green analyses, with the procedures and methodologies defined in Ref. 1 and corresponding c, g) normalized histograms of the nanovesicle distributions, where coincident-, red and green results are highlighted by solid red and dash-dotted green lines, respectively. d, h) difference between the latter curves in the case of c, g), respectively, providing the error ($\varepsilon(R)$) minimization used for optimizing the burst data, as explained in Ref. 1.

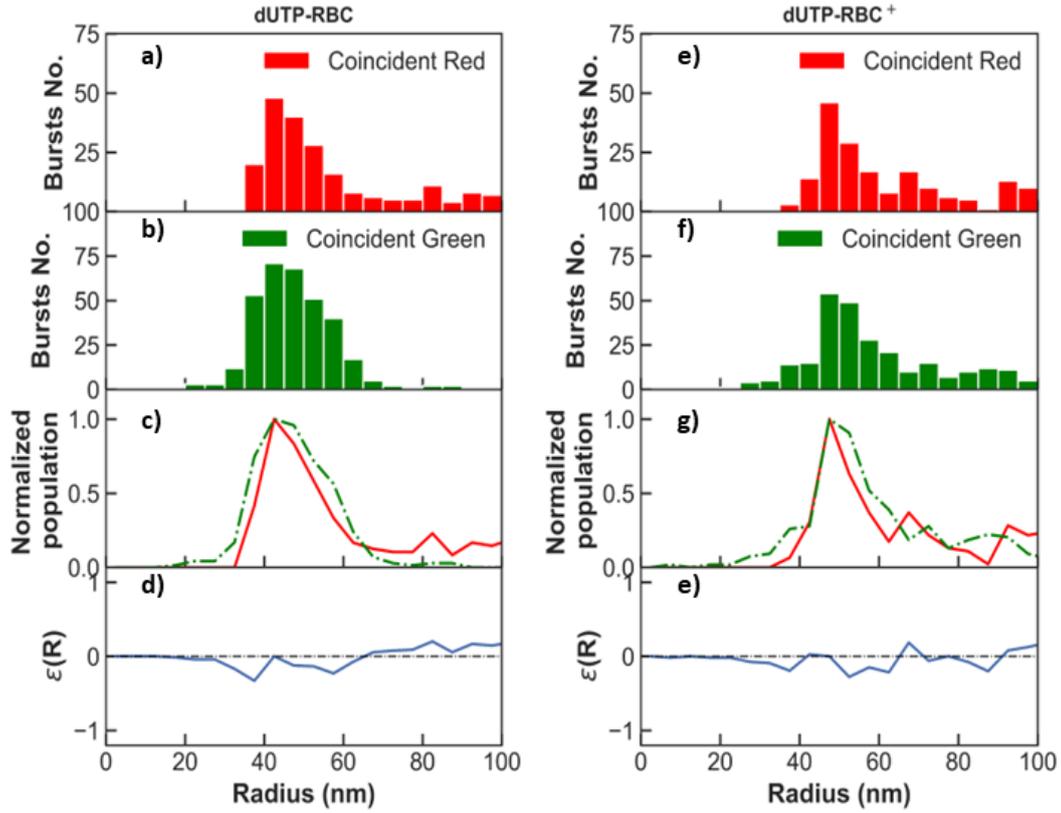

**Figure S8.** Size histograms of loaded nanovesicles obtained from fluorescence experiments for a-c) dUTP-RBC and e-g) dUTP-RBC[+] preparations. In a, e), the histograms are derived from red fluorescence analyses, while in b, f), their result from coincident green analyses are provided, following the procedures and methodologies outlined in Ref. 1. Corresponding c, g) depict normalized histograms of the nanovesicle distributions, derived through coincident-red and coincident-green results, highlighted by solid red and dash-dotted green lines, respectively. In d, h), the difference between the curves presented at c, g), respectively, serving as the basis for error minimization utilized in the optimization of burst data, as detailed in Ref. 1.

| Sample | Ab-RBC | | Ab-RBC[+] | | dUTP-RBC | | dUTP-RBC[+] | |
|---|---|---|---|---|---|---|---|---|
| **Parameters** | Red | Green | Red | Green | Red | Green | Red | Green |
| *F* | 6 | 5 | 8 | 8 | 7 | 5 | 8 | 8 |
| *M* | 28 | 12 | 9 | 9 | 23 | 11 | 7 | 8 |
| $max\{\varepsilon(R)\}$ | 0.36 | | 0.38 | | 0.37 | | 0.28 | |
| $\langle\varepsilon(R)\rangle$ | 0.04 | | 0.04 | | 0.08 | | 0.06 | |

**Table S1.** Summary for the optimized parameters defined according to the data analysis parameters (F: the minimum photon-rate threshold and M: the minimum photon counts per burst) and protocols of Ref. 1 for the red and green fluorescence burst analyses of the four typologies of both cargo and vesicle preparations considered in the study (Ab- /dUTP cargoes and RBC/ RBC[+] preparations). $max\{\varepsilon(R)\}$ and $\langle\varepsilon(R)\rangle$ are the corresponding

values of the maximum and average errors, respectively, for the relevant nanovesicle radius range (R≤100 nm), see also Ref. 1.

Based on the above analysis, the size profile of the loaded nanovesicles were evaluated and the summary of these statistics are provided in Table S2.

| Sample | $R_{max}^{Loaded}$ (nm) | $R_{ave}^{Loaded}$ (nm) | $\sigma_R^{Loaded}$ (nm) |
|---|---|---|---|
| Ab-RBC | 52 | 66 | 15 |
| Ab-RBC+ | 52 | 65 | 15 |
| dUTP-RBC | *(42) [1]* | *(55) [1]* | *(17) [1]* |
| dUTP-RBC+ | *48* | *61* | *17* |

**Table S2.** Summary of the statistics on the size of loaded nanovesicles including, $R_{max}^{Loaded}$: the size of loaded vesicles at the peak of their distribution, $R_{ave}^{Loaded}$: their average size and $\sigma_R^{Loaded}$: corresponding standard deviation.

### 2-4- The lifetimes and brightness assessments

For calibration purposes, the fluorescence measurements were conducted on free green Alexa488-tagged Ab and dUTP cargos in PBS, under the same conditions as the main experiments. Applying the time correlated single photon counting (TCSPC) unit and data acquisition card (Hydraharp, Picoquant) in the DCFM setup, the retrieved lifetime histograms of the green channel for green-tagged cargos within RBC and RBC+ nanovesicles were normalized and compared with the ones for free Ab/dUTP-Alexa488 cargo molecules in solution as shown in Figure 4 in main text and Figure S13 in next section. As apparent from the plots in Figure 4c-d and Figure S13c-d, the lifetimes of the Alexa488 dye inside the nanovesicles (RBC and RBC+), were decreased with respect to the Alexa488 lifetimes of the free (non-encapsulated) Ab- and dUTP- molecules. To account for the observed lifetime shortening due to FRET interaction between Alexa488 and hemoglobulin molecules of RBC, the average count rate (brightness) corresponding to one cargo molecule inside (RBC and RBC+) nanovesicles were determined by including suitable correction factors in the brightness ratios between free cargo-Alexa488 and the ones inside the nanovesicles, inferred from the experimental data. The resulting values are listed in Table S3.

| Brightness ratio | | |
|---|---|---|
| Cargo | RBC nanovesicles | RBC+ nanovesicles |
| Ab-Alexa488 | 0.604 | 0.297 |
| dUTP-Alexa488 | *0.491* | *0.227* |

**Table S3.** Correction factors for the brightness of Ab/dUTP-Alexa488 in RBC and RBC+ nanovesicles, stemming from the modification of the fluorescence lifetimes resulting from the dye encapsulation inside the nanovesicles

due to FRET. The values were determined from systematic comparisons of the normalized lifetime histograms, in measurements of the kind reported in Figure 4c-d at main text and Figure S11 c-d.

## S3- Reference dUTP results and comparative summary with Ab-loading

The main text reports the key results concerning Ab-loading. However, the experimental procedures involved substantial further calibration results and experimental analyses using dUTP-loaded preparations, which were simultaneously processed under the same conditions as the Ab-loaded samples and used as a reference for the Ab-loading results. This section provides summaries of the main results of the corresponding reference studies performed on the dUTP-loaded nanovesicle populations (consistent with the results previously reported in Ref. 1 where the DC-CFB method was developed and applied). Figure S9 represents the size histograms of the whole population of nanovesicles measured by two separate methods including (a-b) AFM and (c-d) red fluorescence obtained from DCFM for dUTP cases in the RBC and extra cleaned RBC$^+$ preparations. The overall comparisons of these measurements along with the ones depicted in Figure 2 are provided in Table 1 at the main text.

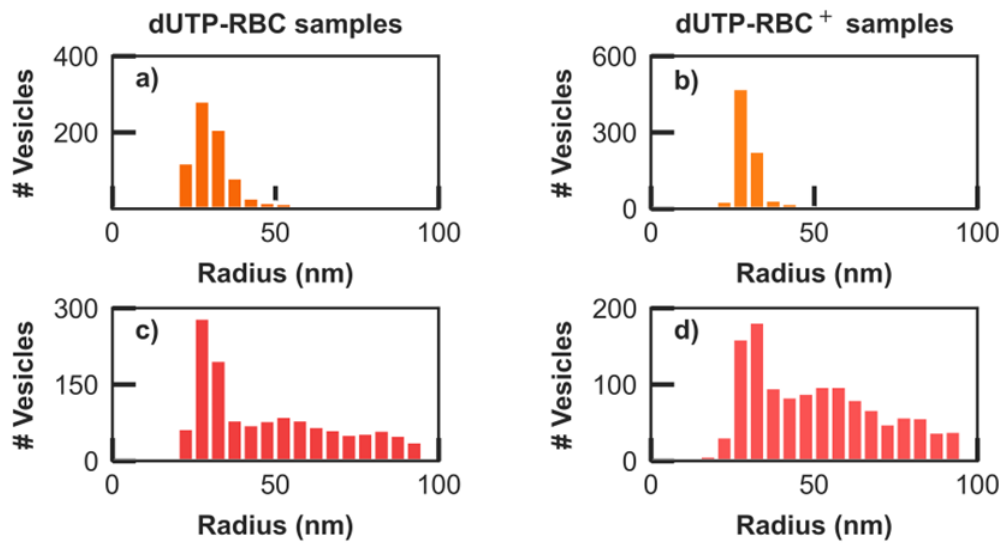

**Figure S9.** The size histograms of the entire (including loaded and unloaded) nanovesicles populations assessed by a-b) AFM and c-d) red fluorescence measurements ($N_{tot}$) in the case of dUTP cargo molecules for: (a and c) RBC, (b and d) RBC$^+$ preparations.

Based on the DC-CFB analysis, the sub-population of the loaded nanovesicles and the loading yield ($\eta(R)$) were assessed while the reference results for dUTP cases are illustrated in Figure S10.

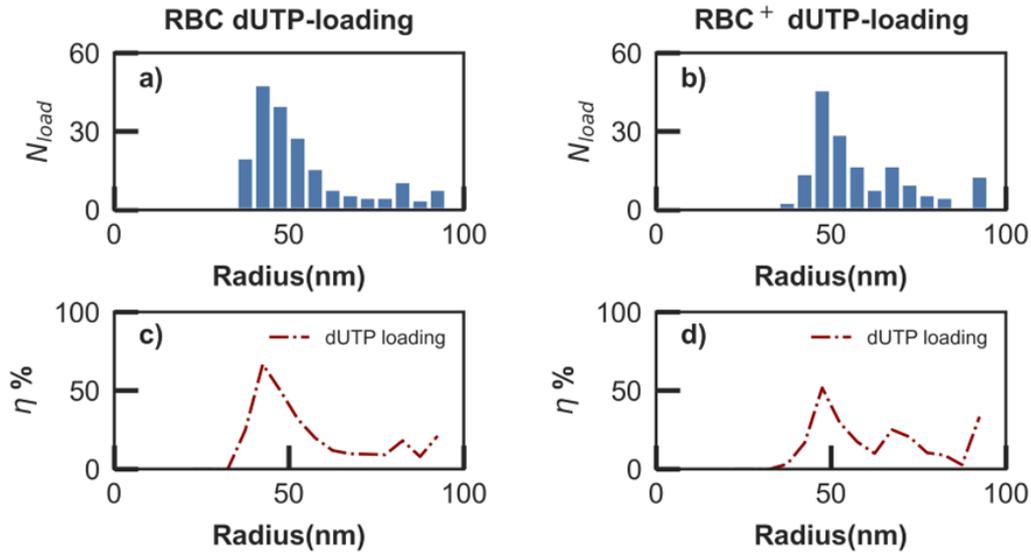

**Figure S10.** Size histograms retrieved from fluorescence measurements for: a-b) the loaded nanovesicle sub-populations ($N_{load}$) and c-d) corresponding loading yield ($\eta(R)$) profile (determined by dual color burst analysis), in the case of dUTP-loaded samples including RBC (a and c) and extra cleaned RBC$^+$ (b and d) preparations.

Moreover, after calibrating the brightness of green-tagged cargos inside the RBC and RBC$^+$ nanovesicles by comparing their lifetimes with the corresponding free cargos in solution as the brightness ratio, listed in Table S3 and shown in Figure S11 c-d.

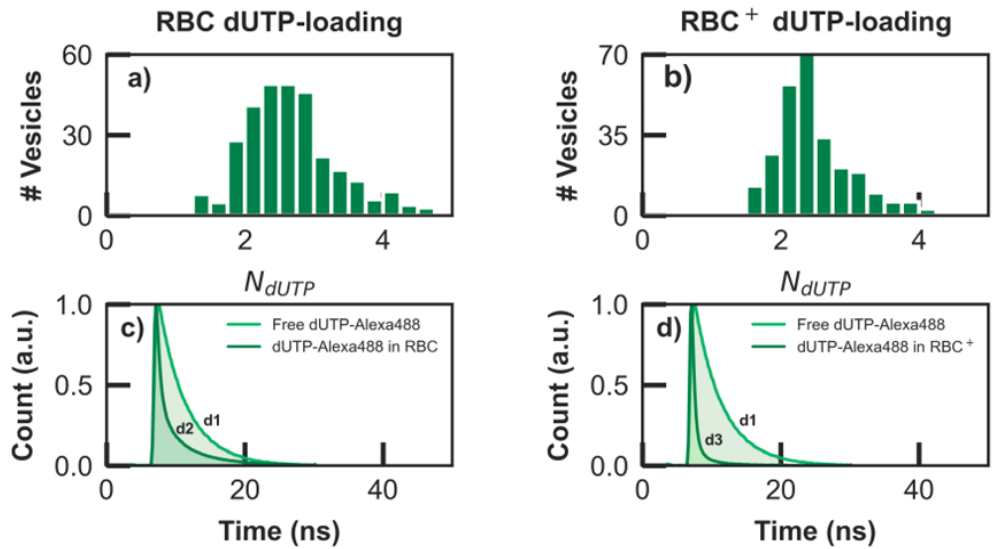

**Figure S11.** The histograms of the number of dUTP cargos ($N_{dUTP}$) loaded in (a) RBC and (b) RBC$^+$ nanovesicle inferred by comparing the normalized lifetime histograms of dUTP-Alexa488 cargos entrapped in (c) RBC ($d_1$) and (d) RBC$^+$ ($d_3$) nanovesicles with the ones measured for free dUTP-Alexa488 molecule ($d_1$).

The histogram of the number of cargos per vesicle has been evaluated and the results for dUTP case ($N_{dUTP}$) are represented in Figure S11 a-b. The lifetime of the Alexa488-tagged dUTP in free solution and encapsulated inside the RBC and RBC[+] nanovesicles are denoted as $d_1$, $d_2$ and $d_3$ in Figure S11 c-d, respectively. To compare the loading yield profiles of Ab-loaded nanovesicles with dUTP-loaded ones, and RBC ones with extra cleaned RBC[+] ones, Figure S12 provides an overview of loading yield distributions considering both dUTP- and Ab-, loaded RBC and RBC[+] nanovesicles. It exhibits that Ab-loading shifted the size of nanovesicles toward slightly larger size compared to dUTP-loaded ones (Fig. S12a). Moreover, the Ab-loading yield was consistently unchanged through extra exosome spin column cleaning (Fig. S12b). However, for dUTP-loaded vesicles, the cleaning procedure had done some size filtering toward slightly larger vesicles comparable to Ab-loaded peak radius (Fig. S12c).

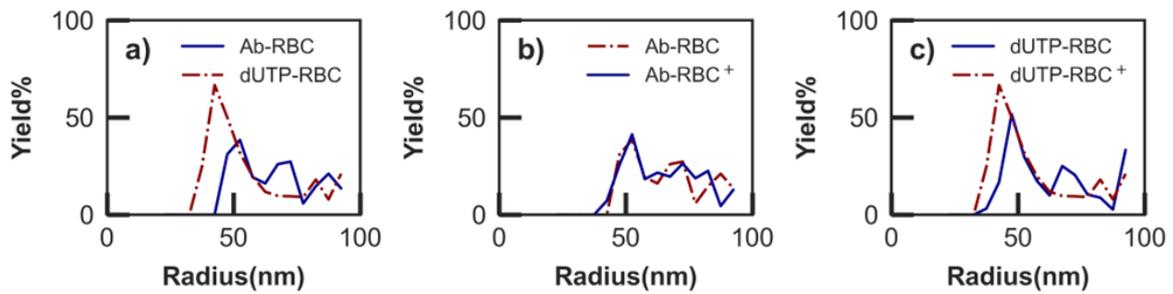

**Figure S12.** Comparisons between the loading yield profiles ($\eta(R)$) for: (a) Ab- versus dUTP- loaded RBC nanovesicles; (b) Ab-loaded RBC versus RBC[+] (extra-cleaned) nanovesicles; (c) dUTP-loaded RBC versus RBC[+] nanovesicles preparations.

In summary, the retrieved two dimensional profile of loaded RBC and RBC[+] nanovesicles, for dUTP case, are illustrated in Figure S13, revealing the most populated size and dUTP-cargos per loaded vesicles, consistent with the results for Ab-loaded ones.

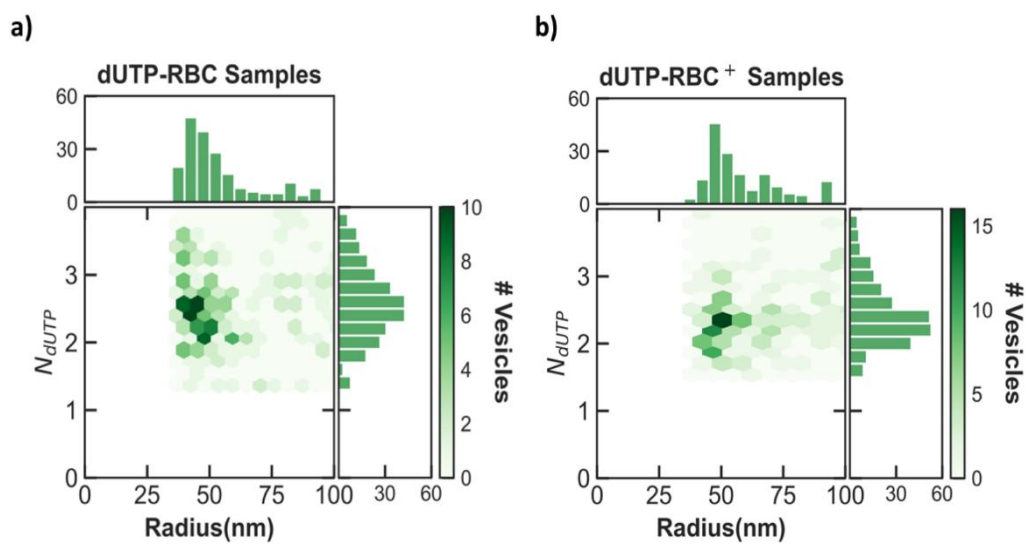

**Figure S13.** Two dimensional profiles of dUTP-loaded a) RBC and b) RBC+ nanovesicles, obtained from dual-color coincident fluorescence burst analysis. The horizontal and vertical subplots depict the size and number-normalized brightness per vesicle for dUTP-cargo molecules, respectively. It demonstrates that loaded nanovesicles are mostly populated at radius of 42 nm for RBC (48 nm for RBC+) and maximum number of 2.25 dUTPs per nanovesicle for both preparations (RBC and RBC$^+$).